\documentclass[9pt,journal,compsoc]{IEEEtran}

\usepackage{graphics}
\usepackage{textcomp}
\newcommand\sbullet[1][.5]{\mathbin{\vcenter{\hbox{\scalebox{#1}{$\bullet$}}}}}

\usepackage{float}
\usepackage{amssymb}
\usepackage{gensymb}
\usepackage{siunitx}

\usepackage{cite}
\usepackage{amsfonts}

\usepackage[font=small,labelfont=bf]{caption}
\usepackage{multicol}
\usepackage{tabularx}
\usepackage{array}
\usepackage[caption=false]{subfig}
\graphicspath{{../}}
\usepackage{booktabs}
\usepackage{pifont}
\usepackage{algorithm,algorithmic,amsmath}
\usepackage[section]{placeins} 

\newcolumntype{t}{>{\ttfamily}l}  
\renewcommand{\baselinestretch}{0.995}
\newcommand*\rot{\rotatebox{90}}
\newcommand*\OK{\ding{51}}
\newcommand{\missingcitation}{[{\color{cyan}\textbf{??}}]}
\newcommand{\abs}[1]{\left\lvert#1\right\rvert}
\usepackage{multirow}
\usepackage{array}
\newcolumntype{P}[1]{>{\centering\arraybackslash}m{#1}}
\usepackage{lipsum}

\usepackage{mathptmx} 

\newcommand{\ignore}[1]{}
\newcommand{\sil}[1]{\textcolor{black}{#1}}
\usepackage{fancyhdr}
\usepackage[normalem]{ulem}
\usepackage[hyphens]{url}
\usepackage{microtype}
\usepackage[svgnames,table]{xcolor}

\usepackage[bookmarks=true,breaklinks=true,letterpaper=true,colorlinks,linkcolor=black,citecolor=black,urlcolor=black]{hyperref}

\newcommand{\myalgo}{SuitAP}
\newcommand{\suitapmech}{Algorithm}
\newcommand{\fixme}[1]{\textbf{\textcolor{red}{FIXME: #1}}}
\newcommand{\answer}[1]{\textbf{\textcolor{blue}{ANSWER: #1}}}
\newcommand{\ajay}[1]{\textbf{\textcolor{orange}{AJAY: #1}}}

\newcommand{\trace}{trace}
\newcommand{\tracein}{data point}
\newcommand{\system}{\textit{L1 + L2 Hybrid Prefetching System}}
\newcommand{\systeml}{\textit{L1 Hybrid Prefetching System}}
\newcommand{\treeprune}{Constructed Tree Pruning}
\newcommand{\classprune}{Training Class Pruning}
\newcommand{\featureprune}{Feature Pruning}
\renewcommand{\arraystretch}{1.2}
\def\code#1{\texttt{#1}}

\usepackage[table]{xcolor}

\usepackage{pifont}

\usepackage{textcomp}
\usepackage{listings}
\usepackage{color}
\usepackage{threeparttable}
\definecolor{dkgreen}{rgb}{0,0.6,0}
\definecolor{gray}{rgb}{0.5,0.5,0.5}
\definecolor{mauve}{rgb}{0.58,0,0.82}

\ifCLASSINFOpdf
   \usepackage[pdftex]{graphicx}
   \DeclareGraphicsExtensions{.pdf,.jpeg,.png}
\else
\fi

\usepackage{array}
\usepackage{booktabs}
\usepackage{multirow}
\usepackage{float}

\usepackage{url}
\usepackage{enumitem}
\usepackage{xcolor}
\usepackage{soul}
\sethlcolor{black}
\makeatletter
\newif\if@blind
\@blindfalse 
\if@blind \sethlcolor{black}\else

\fi

\begin{document}

\title{\huge{Custom Tailored Suite of Random Forests for Prefetcher Adaptation}}

\vspace{-0.1in}
\author{\IEEEauthorblockN{Furkan Eris, Sadullah Canakci, Cansu Demirkiran, and Ajay Joshi}\\ 
  \IEEEauthorblockA{Department of Electrical and Computer Engineering, Boston University\\ 
    \{fe, scanakci, cansu, joshi\}@bu.edu} \vspace{-0.15in}}

%

\IEEEtitleabstractindextext{
\begin{abstract}
To close the gap between memory and processors, and in turn improve performance, there has been an abundance of work in the area of data/instruction prefetcher designs. Prefetchers are deployed in each level of the memory hierarchy, but typically, each prefetcher gets designed without comprehensively accounting for other prefetchers in the system. As a result, these individual prefetcher designs do not always complement each other, and that leads to low average performance gains and/or many negative outliers. In this work, we propose~\myalgo~(\textbf{Suit}e of random forests for \textbf{A}daptation of \textbf{P}refetcher system configuration), which is a hardware prefetcher adapter that uses a suite of random forests to determine at runtime which prefetcher should be ON at each memory level, such that they complement each other. Compared to a design with no prefetchers, using \myalgo~we improve IPC by 46\% on average across traces generated from SPEC2017 suite with $\sim12$KB overhead. Moreover, we also reduce negative outliers using \myalgo. 

\ignore{
Multiple hardware prefetchers are commonly used in today's processors. At runtime, typically all prefetchers are ON, and individual prefetchers are throttled based on the prefetcher accuracy. This baseline configuration can lead
to prefetchers inadvertently sabotaging each other, which in turn leads to a
reduction in application performance. In this work, we propose~\myalgo~(\textbf{Sui}te of \textbf{T}rees for \textbf{A}daptation of \textbf{P}refetcher system configurations), which is a hardware prefetcher adapter that uses a suite of decision trees to determine which prefetchers should be
ON or OFF at runtime. {\color{black} We observe that accuracy is
insufficient as a training metric for hardware adaptation and instead use processor performance (IPC). We design the trees towards the diverse preferences of applications by formulating our algorithm as a multi-label classification algorithm.} We design the hardware for \myalgo~and determine the
Pareto-optimal front for performance and feasibility. Using \myalgo~we improve IPC by 5\% on
average (peak 28\%) across 132 benchmarks with an $\sim11$KB overhead.
}

\end{abstract}

\vspace{-0.05in}
\begin{IEEEkeywords}
Prefetchers, Machine Learning, Hardware Adapters \vspace{-0.05in}
\end{IEEEkeywords}

}

\ignore{
\begin{abstract} When designing general-purpose processors, typically
the hardware settings are statically optimized for aggregate performance
improvements. This optimization leads to lost opportunities as individual
benchmarks benefit from different hardware settings. In this work, we propose to
dynamically configure the hardware prefetchers to improve the overall
Instructions-Per-Cycle (IPC). We approach selecting the best prefetch
configuration as a multi-label classification problem, where a selection of ON
or OFF settings is a ``class''. We use a suite of decision trees, with a tree
per class, to collectively decide the prefetch configuration using runtime
system behavior. We call our machine-learning (ML) based adaptation of
prefetchers as \myalgo~(\textbf{Sui}te of \textbf{T}rees for \textbf{A}daptation
of \textbf{P}refetcher system configurations). We argue that accuracy is
insufficient as a training metric for hardware adaptation and instead use
processor performance. We design the hardware for SuiTAP and determine the
Pareto-optimal front for performance and feasibility. Applying our approach to a
generic out-of-order (OoO) core we see an average of 5\% (peak $28\%$) IPC
improvement across 132 benchmarks. \end{abstract} }

\ignore{ 
\begin{abstract} The memory wall is a well-known challenge in computing
systems. One of the common solutions to overcome the memory wall has been
prefetching. {\color{black} To improve prefetching, researchers have} focused on
implementing multiple prefetchers (hereafter, referred to as ``Hybrid
Prefetching System'') or increasing the complexity of existing prefetchers.
{\color{black}Both} approaches lead to an increase in {\color{black}the number
of} configuration choices for prefetchers (hereafter, referred to as ``Prefetch
Configuration''). While {\color{black}the choice of prefetch configuration} can
be statically optimized for aggregate performance improvements, there are lost
opportunities where individual benchmarks prefer alternative prefetch
configurations. In this work, we observe that by dynamically configuring the
hybrid prefetching system overall Instructions-Per-Cycle (IPC) can be increased.
We approach selecting the best prefetch configuration as a multi-label
classification problem, where a selection of ON or OFF setting {\color{black}of
each} individual prefetchers is a ``class''. We propose using a suite of
decision trees, with one tree per class. These trees collectively decide on
which prefetch configuration to use for the currently observed system behavior.
We call our {\color{black}machine-learning (ML) based adaptation of prefetchers}
as \myalgo~(\textbf{Sui}te of \textbf{T}rees for \textbf{A}daptation of hardware
\textbf{P}refetch configurations). We show that our method can improve a hybrid
prefetching system with minimal overhead in hardware. {\color{black}Applying our
approach to} a generic out-of-order (OoO) core, based on offline analysis of our
dataset we see an average of 5\% (peak value $28\%$) IPC improvement across 132
benchmarks from six benchmark suites. \end{abstract}}


\maketitle

\IEEEdisplaynontitleabstractindextext

\IEEEpeerreviewmaketitle

\section{INTRODUCTION}
\label{sec:intr}
The memory wall is a well-known problem in computer architecture \cite{wulf1995hitting}. One method used to combat the memory wall is data/instruction prefetching. To this end, computer architects have developed many different hardware prefetchers \cite{falsafi2014primer}. Today's processors consists of multiple prefetchers at each level in the memory hierarchy \cite{bios2010kernel, guide2011intel}. {\color{black} The use of these prefetchers can lead to high average IPC gain, but can have some applications losing performance. Our experiments show that we have large negative outliers even in the current state-of-the-art prefetchers such as Bingo \cite{bakhshalipour2019bingo} and IPCP \cite{Pakalapati2020ipcp}, when they are used with other prefetchers. This case is exacerbated when individual prefetchers are designed without accounting for the other prefetchers in the system.} Multiple challenges arise from the interactions between the multiple prefetchers -- 1)
\ignore{
The hardware prefetching system in today's processors consists of multiple prefetchers, and this prefetcher system follows a pre-determined protocol for prefetcher priority, prefetcher throttling, etc. to regulate the interactions between prefetchers. Generally, this protocol is tuned for the common case and this tuning is done at design time in both AMD\ignore{\footnote{AMD, the AMD Arrow logo and combinations thereof are trademarks of Advanced Micro Devices, Inc.}} and Intel CPUs \cite{bios2010kernel,guide2011intel}.} 
\ignore{
{\color{black}Only tuning for the common case can cause complex interactions between prefetchers which can lead to IPC losses.} Despite this tuning,}
{\color{black}Each prefetcher tracks a specific type of traffic. For example, a stride prefetcher tracks strided accesses for a single stride, and it uses thresholds tuned for the average strided sequences. For applications that do not have strided accesses, this stride prefetcher may be suboptimal, leading to cache thrashing. In a prefetching system consisting of multiple prefetchers, this issue is more pronounced because the architectural resources are shared among all prefetchers;}
\ignore{
\textbf{Duplicate Coverage -} {\color{black} Given that prefetchers in a prefetcher system are trained independently to track a specific type of traffic, multiple prefetchers may trigger prefetch requests for the same memory patterns. This leads to wasted bandwidth and increased power consumption.} \break
}
{\color{black} and 2) Different prefetchers latch onto memory access patterns at different speeds, and so a prefetcher's behavior over time can be affected by the traffic of the other prefetchers. These differences in temporal behavior can cause faulty synchronization among prefetchers. Variations in the accuracy of each prefetcher and faulty synchronization can lead to a drop in performance.}
\ignore{
\begin{figure}[t] \centering
  \includegraphics[width=0.95\columnwidth]{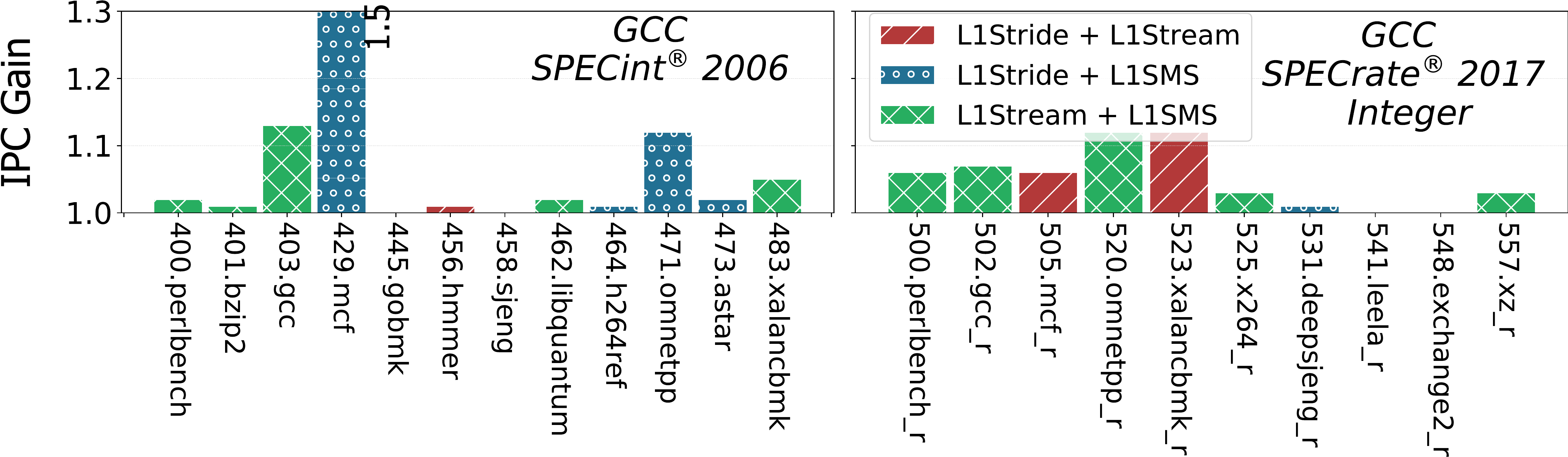}
  \caption{IPC Gain (normalized to ???) for different benchmarks running on an
  x86 processor. For each benchmark, we only show the bar for the prefetcher
  system configuration that gives the highest IPC gain.}
  \label{fig:motivation}
\end{figure}
Essentially, prefetchers compete for resources, and at times, sabotage each
other. One way to address this problem, one could switch ON a subset
of prefetchers by identifying which prefetchers are best for a given program
phase. In Figure~\ref{fig:motivation}, we present evidence to back up this
claim. We compile SPEC CPU\textsuperscript \textregistered~2006 and SPEC
CPU\textsuperscript \textregistered~2017 using \textit{gcc} (Section
\ref{sec:methodology}).\footnote{SPEC\textsuperscript \textregistered, SPEC
CPU\textsuperscript \textregistered, SPECint\textsuperscript \textregistered,
SPECfp\textsuperscript \textregistered, SPECrate\textsuperscript
\textregistered, SPECjbb\textsuperscript \textregistered, and
SPECweb\textsuperscript \textregistered~are registered trademarks of the
Standard Performance Evaluation Corporation. More information can be found in
the given citations \cite{SPEC2017, SPEC2006, Legacyweb, Contempspec}.} for an 
x86 target system. We use a system with three prefetchers: (i) L1Stride, (ii) 
L1SMS, and (iii) L1Stream. For each benchmark we determined the IPC for each 
one of the eight prefetcher system configurations (PSCs), corresponding to the 
ON/OFF combinations of the three prefetchers. We normalized these IPC values to
the IPC where all three prefetchers are ON to determine the IPC gain. For each benchmark we show the PSC that gives us the highest IPC gain. From Figure~\ref{fig:motivation}, we can see
that the PSC that gives the highest IPC gain varies across benchmarks. To
achieve the highest possible gain, we propose using Machine Learning (ML) to
determine the best PSC at runtime. The contributions of our work are: \break
}
\break
Essentially, prefetchers compete for resources, and at times, sabotage each other. To validate our argument, we generate traces from SPEC2017 benchmarks and {\color{black}run these traces using different prefetchers at various memory levels. We observe that the performance swing between the best and worst prefetcher system configuration (PSC)\footnote{A PSC specifies which prefetcher is switched ON at each level of the memory in the system. A PSC is denoted as prefetcher-in-L1I\$)-(prefetcher-in-L1D\$)-(prefetcher-in-L2\$)-(prefetcher--in-LL\$). The prefetchers available at each level are provided in Figure \ref{fig:sysarchitecture}.} can be very large (up to 642\%).} One way to address this problem is one could switch ON a subset of prefetchers that complement each other and are best for a given program phase. 

To achieve the highest possible gain in application performance, we propose using machine learning (ML) to determine the best PSC at runtime. Contrary to the prior work \cite{liao2009machine}, in our approach, we train our ML model to increase the overall processor IPC instead of training the model to improve the accuracy. We adapt a multi-label classification approach for prefetcher adaptation, where we leverage the implicit ranking among PSCs for each application to train our ML model to catch performance outliers. We train a unique random forest per PSC to create a suite of random forests. \ignore{ A more detailed discussion about our choice of multi-label classification and suite of random forests is provided in Section~\ref{sec:suitap}.} {\color{black}We use hardware-invariant events as our features}. In particular, we choose events that are not affected by the choice of PSCs. We design our ML model with the hardware overhead in mind and aim to maximize IPC with minimum overhead. 
The contributions of our work are: \break
$\bullet$ \textbf{Design - }We design \myalgo, a hardware adapter, which uses a suite of random forests to determine the PSC at runtime. \myalgo~is non-invasive and complements any prefetcher design or heuristic. {\color{black}We leverage hardware-invariant events to train \myalgo~to make it agnostic of the changing processor conditions.} \break
$\bullet$ \textbf{Evaluation - }{\color{black}We train \myalgo~to maximize processor performance instead of accuracy. We use only 10\% of the data for training to prevent overfitting and design \myalgo~with a low hardware overhead ($\sim12$KB). We ensure \myalgo~reduces negative performance outliers. For the traces generated from SPEC2017 benchmarks, \myalgo~achieves an average performance gain of over 46\% compared to a system with no prefetching.}

\ignore{ 

\myalgo~reduces the performance loss in negative outliers by 18\% compared to Bingo and 23.5\% compared to IPCP. 

\break $\bullet$ \textbf{Design - }We design \myalgo, a hardware
adapter which utilizes ML for hardware prefetcher adaptation. \myalgo~employs a
suite of binary trees to perform multi-label classification for many different
hybrid prefetching systems. Our method is non-invasive and does not need
modifications to the existing prefetchers. We explore the design space of
\myalgo~to determine the Pareto optimal front along which we trade off
performance with hardware overhead. \break $\bullet$ \textbf{Implementation -
}We use pruning methods to reduce the number of {\color{black}
{\color{black}hybrid prefetcher setting} choices, the number of {\color{black}hybrid
prefetcher setting}s used for training \myalgo, and the number of performance
metrics that \myalgo~uses for training. We construct an auto-optimization
framework,} where trees can {\color{black}auto-optimize hardware resources.}
\break $\bullet$ \textbf{Evaluation - }We test and train \myalgo~using a
{\color{black}large industry-standard dataset made up of 6 benchmark suites
containing 132 benchmarks, or in total 2051 \tracein s constructed from these
benchmarks.} We construct \myalgo~with robustness guarantees against changing
hardware and program environments at runtime. We show performance improvements
(up to $28\%$) across almost all benchmarks with very few negative outliers
($<1\%$ of benchmarks) based on offline analysis. }
\ignore{
Heuristics are used
to statically decide the order in which prefetch requests from different
prefetchers are issued \cite{dhodapkar2002managing, kang2013hardware} or
statically select one prefetcher to issue requests. Heuristics use commonsense
parameters (e.g., L1 cache misses) to tune interactions between prefetchers at
design-time. These parameters may not be correct or may not work at runtime. In
fact, heuristics limit the adaptability of reconfiguration schemes and may even
lead to performance degradation by causing late prefetches, over-consuming cache
bandwidth, and sub-optimal coverage/accuracy. Our method which employs machine
learning (ML) is superior to heuristic-based techniques
\cite{rahmani2018spectr,pothukuchi2016using}. ML enables higher performance
bounds as ML algorithms are capable of handling a significantly larger number of
parameters. ML can also learn unintuitive interactions between parameters. In
this work, we use ML to adapt hardware prefetchers settings at runtime.}
\ignore{
\footnote{SPEC\textsuperscript \textregistered, SPEC
CPU\textsuperscript \textregistered, SPECint\textsuperscript \textregistered,
SPECfp\textsuperscript \textregistered, SPECrate\textsuperscript
\textregistered, SPECjbb\textsuperscript \textregistered, and
SPECweb\textsuperscript \textregistered~are registered trademarks of the
Standard Performance Evaluation Corporation. More information can be found in
the given citations \cite{SPEC2017, SPEC2006, Legacyweb, Contempspec}.}}

\section{Related Work}
\label{sec:related}

\ignore{ 
Prior work has utilized hardware adaptation schemes for improving the
IPC and the power efficiency of the overall system \cite{dubach2010predictive,
subramanian2013mise, bitirgen2008coordinated}; providing performance
\cite{subramanian2015application, subramanian2013mise}, memory
\cite{ebrahimi2010fairness, subramanian2013mise} and network
\cite{das2009application} fairness in the management of programs; monitoring,
controlling and predicting the thermal behavior \cite{bartolini2011distributed},
and adjusting frequency, energy, and power usage of a system
\cite{cochran2011pack, su2014ppep, wang2011adaptive, bartolini2011distributed}.
Specifically, in the area of prefetcher adaptation, prefetcher throttling
methods are close to our work. In prefetcher throttling, depending on prefetcher
metrics, such as accuracy or misses per demand, prefetchers change
aggressiveness in order to increase IPC \cite{srinath2007feedback,
ebrahimi2009coordinated, ebrahimi2011prefetch}. Besides prefetcher throttling,
Liao \textit{et al.} \cite{liao2009machine} have investigated the accuracy of
various ML techniques for predicting prefetch configurations, Jimenez \textit{et
al.} \cite{jimenez2012making} utilized heuristics to increase performance in an
IBM POWER7, and Kang \textit{et al.} \cite{kang2013hardware} proposed to turn
all prefetching ON/OFF. }
Broadly, heuristic methods based on human intuition have been used in the past for designing hardware adapters for prefetcher systems\cite{ebrahimi2009coordinated}. Heuristic-based approaches are typically comprised of simple rules that designers have found based on intuitions gained from experimentation. While these heuristic methods improve processor performance, they are not optimal and leave performance on the table. 

\ignore{
\textbf{(Column 2)} Some of the prior work utilize unrealistic baselines where there are no prefetchers or only a single prefetcher \cite{bhatia2019perceptron, jimenez2012making}. \textbf{(Column 5)} Some of the prior work have small datasets and do not show evidence for large-scale industrial datasets \cite{bhatia2019perceptron, hiebel2019machine,jimenez2012making, rahman2015maximizing}. \textbf{(Column 6)} Some of the prior work, while providing some level of adaptation, are largely rule-based and do not provide runtime adaptation \cite{jimenez2012making}. \textbf{(Columns 7-8)} Some of the prior work are not designed with hardware adaptation in mind and only wish to gain some insights from data or are very highly data-dependent \cite{liao2009machine, jimenez2012making, hiebel2019machine}.}
\ignore{
In Table \ref{tab:prior}, we compare various prefetcher adaptation methods with our work. Note that because of the wide variety of systems and datasets used for evaluation in these works, a direct head-to-head comparison of \myalgo~with prior work using accuracy and hardware overhead metrics is difficult. So we make qualitative comparisons.}
\ignore{
Some of the prior work utilize unrealistic baselines where there is only a single prefetcher \cite{bhatia2019perceptron, jimenez2012making}. Using a prefetcher system with a limited number of prefetchers limits the scope for performance improvement via prefetching. In a highly-tuned modern processor, there are several prefetchers that work together. {\color{black}We utilize a modern OoO core with multiple prefetcher options at each memory level.}}
Recently, ML methods have been gaining traction in place of heuristic methods for prefetcher adaptation \cite{bhatia2019perceptron, liao2009machine, jimenez2012making, hiebel2019machine}. These methods are capable of extracting the non-intuitive interactions between the different prefetchers. Prior ML methods on prefetcher adaptation configure or train the adapter using the static preset PSC and with small datasets \cite{liao2009machine, jimenez2012making,  bhatia2019perceptron}. Using an ML model trained using only the static preset PSC would make sense if the prefetcher system always stays in the static preset PSC at runtime. However, the PSC changes over time and it could affect the characteristics of the E-PTI\footnote{E-PTI corresponds to the average number of occurrences of a hardware event per thousand instructions. We use E-PTI as the features of our ML model.} values collected at runtime. These characteristics could be different from the characteristics of the E-PTI values corresponding to the static preset PSC. If we use an ML model trained only based on the E-PTI values of the static preset PSC, it will lead to a sub-optimal choice of PSC at runtime. When training the ML model, we need to account for the fact that a prefetcher system can be in any one of its {\color{black}$N_{psc}$ possible configurations.} To address this concern we train our ML model using hardware-invariant events as features.

We observe a wide variation in the complexity of the ML algorithms used in prior work. Some of the algorithms are simple and show good results because they either use small datasets or use datasets that do not accurately portray the runtime environment. As a result, these algorithms can not achieve good accuracy at runtime \cite{liao2009machine, jimenez2012making}. Other algorithms, such as neural networks, are too complex and do not scale well with the size of the dataset \cite{bhatia2019perceptron}. Moreover, some prior works focus on hardware adaptation only from the perspective of accuracy without worrying about the hardware implementation \cite{liao2009machine, jimenez2012making, hiebel2019machine}. 

In our work, we jointly account for accuracy and hardware when designing \myalgo. \myalgo~is complex enough to provide good accuracy on {\color{black}a wide variety of micro-behavior.} At the same time, \myalgo~is not too complex to implement in hardware and scales well with the number of prefetchers and the size/complexity of the dataset.

\section{\myalgo~Design}
\label{sec:suitap}
\ignore{
In this section, we first provide a broader system-level overview of \myalgo, followed by detailed description of the algorithm and microarchitecture of \myalgo.}

\subsection{\myalgo~System Level Overview}
\label{sec:suitap_overview}
\begin{figure}[t]
  \centering
  \includegraphics[width=0.9\columnwidth]{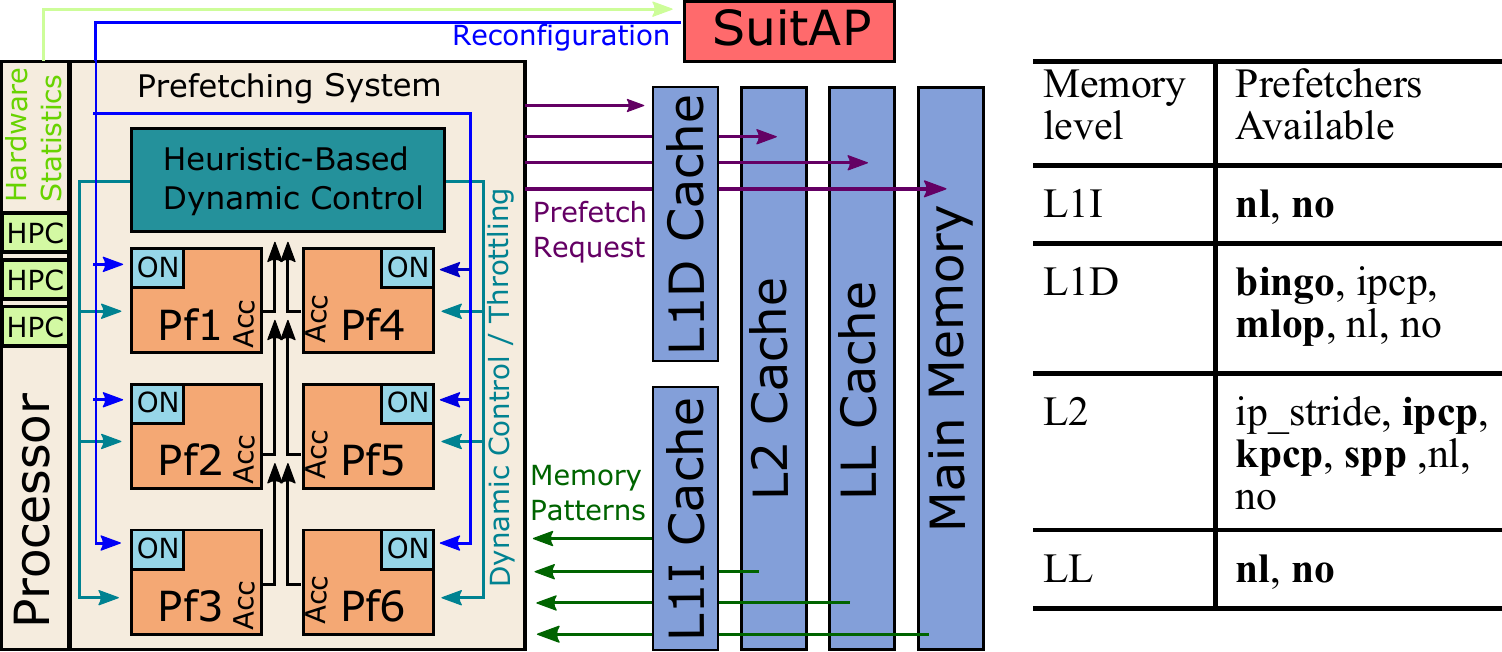}
  \caption{System-level view of an example prefetching system that uses \myalgo, and prefetchers that are available to \myalgo~at each memory level. Prefetchers available to \myalgo~after reducing the number of PSCs are shown in bold. Here Pf = prefetchers, nl = next line prefetcher, mlop = multi-lookahead offset prefetcher, ipcp = instruction pointer classifer-based spatial prefetcher, kpcp = kill-the-PC prefetcher, bingo = bingo prefetcher, spp = signature path prefetcher, and no = no prefetching.}
  \label{fig:sysarchitecture}
\end{figure}

In Figure \ref{fig:sysarchitecture}, we show the system-level design of an example prefetching system that uses \myalgo. The prefetchers track memory access patterns and send requests to prefetch data from main memory to LL\$, from LL\$ to L2\$, and from L2\$ to L1\$. These prefetchers can sometimes act overly aggressively, and can adversely affect each other, in turn leading to loss of application performance. There are many heuristics-based mechanisms that use accuracy of the prefetchers or memory bandwidth utilization to throttle prefetchers in such adverse scenarios~\cite{ebrahimi2009coordinated}.
\ignore{
These mechanisms need to be very simple in order to act at the cycle-level and are generally highly tuned for the given prefetcher system.}
\myalgo~works as a meta-controller and complements these heuristics-based throttling mechanisms. At runtime, it periodically updates the PSC i.e. it sets which prefetcher should be ON and which should be OFF at each level in the memory hierarchy, and allows each prefetcher to continue to use its associated heuristic-based throttling mechanism. To update the PSC, \myalgo~uses an ML model with the E-PTI values of specific events as inputs to determine the next PSC. Effectively, {\color{black} throttling mechanisms are used in prefetchers to regulate the} short-term behavior of the prefetchers, while \myalgo~controls the longer-term system-level behavior using a more complicated ML-based approach.

\subsection{\myalgo~\suitapmech}
\label{sec:suitap-mech}

\ignore{
\begin{table}[]
  \centering{
  \scriptsize{
  \caption{\textbf{Prefetch Preferences -} Best performing classes i.e.
PSC for different benchmarks for a system with $N_{psc}=8$.
\ignore{The example
  architecture contains three prefetchers, thus,
eight
  classes exist.}}
    \label{tab:app_example}
\begin{tabular}{l|l|l|l|l|l|l|l|l}
\hline
\textbf{Class}  & \textbf{0} & \textbf{1} & \textbf{2} &
\textbf{3} & \textbf{4} & \textbf{5} &
\textbf{6} & \textbf{7} \\ \hline \hline
  \textit{548.exchange2\_r} &  \OK               &     \OK
  &         \OK         &      \OK            &         \OK         &
\OK       &          \OK        &     \OK             \\ \hline
  \textit{523.xalancbmk\_r} & \OK                &     \OK
 &                 &                 &                 &                 &
          &                 \\ \hline
  \textit{525.x264\_r} &   \OK              &                 &
\OK           &                 &      \OK           &      \OK           &
    \OK       &                 \\ \hline
  \textit{401.bzip2} &                 &                 &
            &                &                 &                &
&     \OK            \\ \hline
\end{tabular}
}
}
\end{table}
}
\ignore{
\begin{table}[b]
  \centering{
  \scriptsize{
  \caption{\textbf{Prefetch Preferences -} Best performing classes i.e.
prefetch
configurations for different benchmarks for a system with $N_{pf}=3$.
\ignore{The example
  architecture contains three prefetchers, thus,
eight
  classes exist.}}
    \label{tab:app_example}
\begin{tabular}{|l|l|l|l|l|l|l|l|l|}
\hline
\textbf{Class}  & \textbf{0} & \textbf{1} & \textbf{2} &
\textbf{3} & \textbf{4} & \textbf{5} &
\textbf{6} & \textbf{7} \\ \hline
  \textit{548.exchange2\_r} &  1.0               &     1.001
  &         1.0         &      1.0            &         1.004         &
1.003      &          1.004        &     1.002             \\ \hline
  \textit{523.xalancbmk\_r} & 1.0                &     1.002
 &                 &                 &                 &                 &
          &                 \\ \hline
  \textit{525.x264\_r} &   1.0              &                 &
1.003           &                 &      1.005          &      1.004          &
    1.0008      &                 \\ \hline
  \textit{401.bzip2} &                 &                 &
            &                &                 &                &
&     1.5            \\ \hline
\end{tabular}
}
}
\end{table}

\begin{table}[b]
  \centering{
  \scriptsize{
  \caption{\textbf{Prefetch Preferences -} Best performing classes i.e.
prefetch
configurations for different benchmarks for a system with $N_{psc}=8$.
\ignore{The example
  architecture contains three prefetchers, thus,
eight
  classes exist.}}
    \label{tab:app_example}
\begin{tabular}{|l|l|l|l|l|l|l|l|l|}
\hline
\textbf{Class}  & \textbf{0} & \textbf{1} & \textbf{2} &
\textbf{3} & \textbf{4} & \textbf{5} &
\textbf{6} & \textbf{7} \\ \hline
  \textit{548.exchange2\_r} &  0.996               &     0.997
  &         0.996         &      0.996            &         1.0         &
0.999     &          1.0        &     0.998             \\ \hline
  \textit{523.xalancbmk\_r} & 0.998                &     1.0
 &                 &                 &                 &                 &
          &                 \\ \hline
  \textit{525.x264\_r} &   0.995              &                 &
0..98           &                 &      1.0          &      0.999         &
    0.995      &                 \\ \hline
  \textit{401.bzip2} &                 &                 &
            &                &                 &                &
&     1.0            \\ \hline
\end{tabular}
}
}
\end{table}
}
\ignore{
\subsubsection{Multi-Label vs. Binary Classification}
\label{sec:multi-vs-binary}
Our analysis shows that multiple PSCs are sometimes suitable for a \trace. {\color{black}For example for the \textit{607.cactuBSSN\_s-2421B} trace from SPEC2017, only a few PSCs are suitable. On the other hand, for the \textit{638.imagick\_s-10316B} trace all PSCs are suitable.} Using binary classification, which tries to find the best PSC, can lead to a sub-optimal choice of the PSC. Multi-label classification on the other hand is aware of the multiple suitable PSCs and results in a better choice of PSC. So we use multi-label classification in our ML model.
}

\subsubsection{Suite of Random Forests vs. Monolithic Tree}
\label{sec:monovsmulti}
\begin{figure}[]
  \centering
    \subfloat[Suite of Random Forests]{
    \includegraphics[clip,width=0.65\columnwidth]{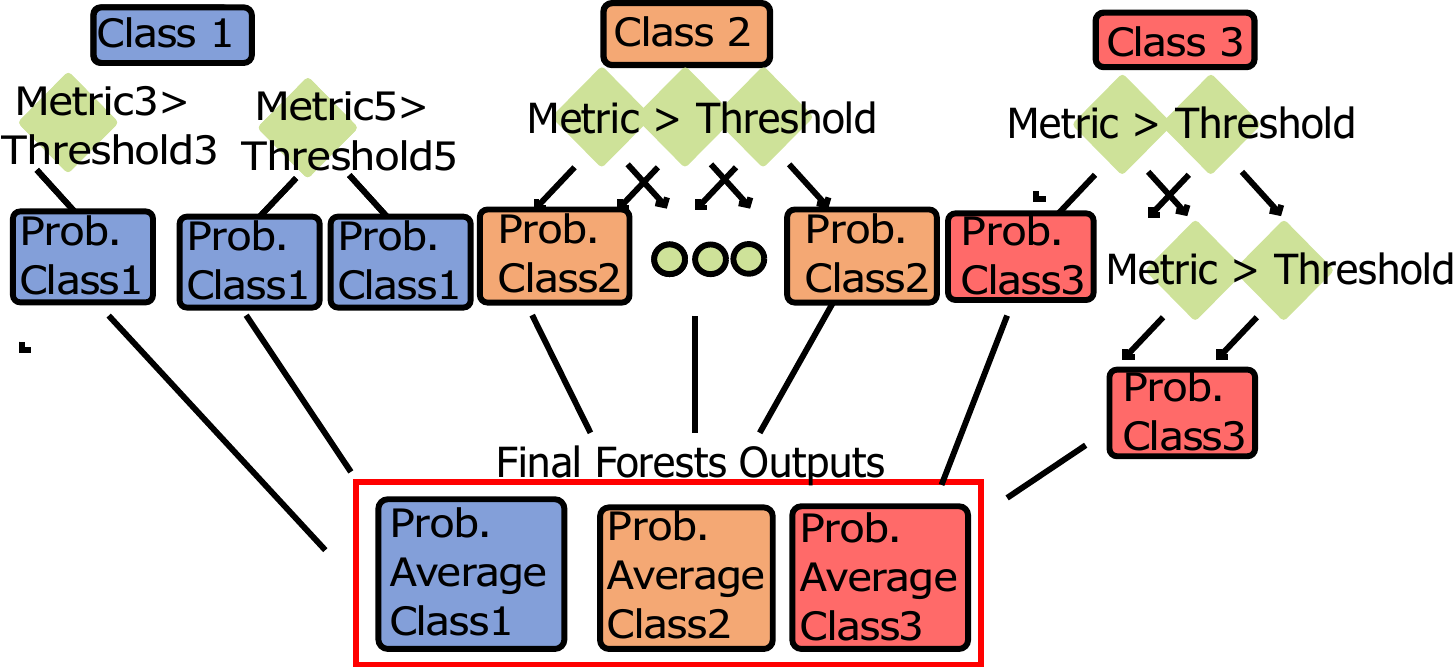}
  }
  \subfloat[Monolithic Tree]{
    \includegraphics[clip,width=0.32\columnwidth]{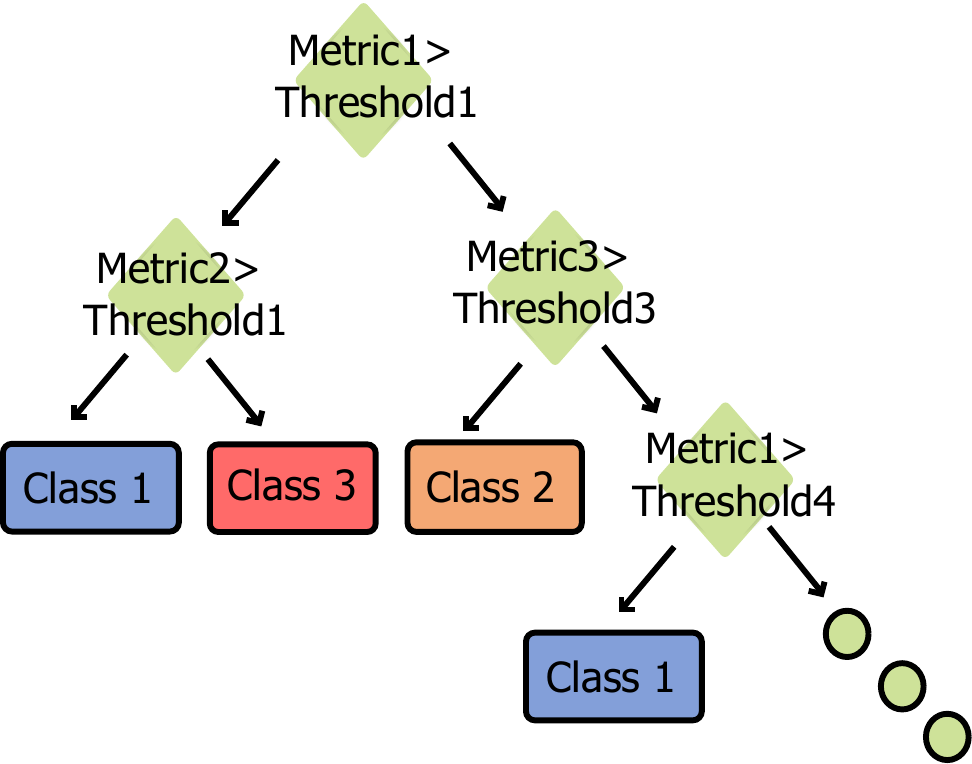}
  }
  \caption{Suite of Random Forests and Monolithic Tree - Each tree in each random forest outputs a PSC probability that is averaged per forest. The Monolithic tree outputs a single PSC choice.}
  \label{fig:method_compare}
  \vspace{-0.2in}
\end{figure}

We use multi-label classification for our ML model and implement it using decision trees due to their simpler implementation. We consider the following two types of decision trees (see Figure \ref{fig:method_compare}): (i) Monolithic tree -- Here we train a single tree for all classes. We split the tree such that we maximize the accuracy across all classes in unison instead of maximizing the accuracy of each class separately; and (ii) Suite of random forests -- Here we train a custom random forest per class i.e. per PSC. As a result, we allow each forest to optimally split at locations that are unique to that class. When using a monolithic tree, at runtime, we traverse down the tree to a leaf corresponding to the next best PSC. 
\ignore{
Liao \textit{et al.} \cite{liao2009machine} proposed using such a monolithic tree for prefetcher adaptation in data centers.}
In a suite of random forests, we have multiple trees per forest and leaves of a tree specify the probability that the PSC associated with that forest is the next best configuration. For each forest, we calculate the average of the probability values obtained from all the trees in the forest, and then choose the PSC with the highest probability. We choose suite of random forests over monolithic tree because it provides better accuracy.

In our classification problem, for a given application phase, there is an implicit ranking among the PSC choices based on the IPC of each PSC i.e. each class. This ranking shows that sometimes an application phase is indifferent to the PSC, and at other times, it is very sensitive with as much as $642\%$ change in IPC. 
Thus, we would like to note that the accuracy of classification does not always translate to an improvement in overall IPC. Even when classified perfectly, the application phases that are indifferent to PSC do not see a change in the overall IPC. For the application phases that are very sensitive to PSC, we can have large IPC gains or losses.
\ignore{In other words, an ML model may correctly choose the PSC $90\%$ of the time (for the indifferent application phases), but for the remaining $10\%$ (for the application phases that are sensitive) can still lower the overall IPC gain (or even cause overall IPC loss) when classified incorrectly.}


\ignore{
\subsubsection{\color{black}Suite of Random Forests vs. Neural Networks} Several prior works have used neural-network based algorithms for runtime hardware adaptation \cite{bhatia2019perceptron, yin2020experiences}. While these algorithms are powerful, {\color{black}NN-based algorithms are generally very large.} For example, LSTMs typically have millions of parameters, cannot fit in on-chip cache, and have large memory access overheads. Thus, neural-network based algorithms are not suitable for \myalgo.
\ignore{In fact, some recent works have suggested using LSTMs to help human designers implement heuristic rule-based approaches \cite{yin2020experiences}. These authors conclude that implementing the LSTM in hardware is not feasible and require human-designers to infer offline meaning.}
}
\subsubsection{Training \myalgo}
\label{sec:training_details}

We train \myalgo~using a diverse set of traces generated from SPEC2017 benchmarks. Consider the case where we have a single prefetcher, $Pf$, {\color{black}at only one memory level}. Here $N_{psc} = 2$ with $Pf$=OFF or $Pf$=ON as the two PSCs. For two consecutive instruction windows, we will have $N_{psc}^2 = 4$ possible scenarios: (i) $Pf$=OFF for both windows, (ii) $Pf$=ON for the first window then OFF for the second, (iii) $Pf$=OFF for the first window then ON for the second, and (iv) $Pf$=ON for both windows. With $N$ number of instruction windows and $N_{app}$ number of applications, the number of different possible application behavior scenarios will then be $N_{app} * N_{psc}^N$. When $N$ increases, the number of different scenarios will increase exponentially. Accounting for each unique scenario during training is not feasible.

\ignore{
\begin{table}[]
  \centering
  \caption{\textbf{Prefetchers Available at Each Memory Level} - prefetchers used by \myalgo~after reducing the number of PSCs shown in bold.}
  \label{tab:prefetchers}
\scriptsize
\renewcommand{\arraystretch}{1.0}
\centering
\center{
\begin{tabular}{l|P{5cm}}
\hline
\textbf{Memory Level} & \textbf{Prefetchers Available}\\
\hline \hline
$L1I$ & \textbf{nl}, \textbf{no}.\\
\hline
$L1D$ & \textbf{bingo}, ipcp, \textbf{mlop}, nl, no. \\
\hline
$L2$ & ip\_stride, \textbf{ipcp}, \textbf{kpcp}, \textbf{spp}, nl, no\\
\hline
$LL$ &  \textbf{nl}, \textbf{no}.
\\
\hline
\end{tabular}
}
\vspace{-0.1in}
\end{table}
}

\begin{table}[t]
  \centering
  \caption{Hardware-level events used by \myalgo.}
  \vspace{-0.1in}
  \label{tab:feature_set}
\scriptsize
\renewcommand{\arraystretch}{1.0}
\centering
\center{
\begin{tabular}{l|P{5cm}}
\hline
\textbf{Hardware Event} & \textbf{Properties}\\
\hline \hline
$RQ\_ROW\_BUFFER\_HIT$ & Number of buffer hits in the read queue buffer.\\
\hline
$LL\$\_LOAD\_HIT$ & Number of LL\$ hits on load. \\
\hline
$BRANCH\_DIRECT\_JUMP$ & Number of direct jumps on branch. \\
\hline
$L1I\$\_LOAD\_MISS$ &  Number of L1I\$ load misses. \\
\hline
$L2\$\_PAGES\_PREFETCHED$ & Number of unique L2\$ pages prefetched. \\
\hline
$BRANCH\_CONDITIONAL$ & Number of conditional branches.
\\
\hline
\end{tabular}
\vspace{-0.05in}
}
\end{table}

\begin{table}[h]
  \caption{{PSCs used by \myalgo -} The ticks
  indicate that the trace achieves top ten performance for the corresponding PSC.}
  \label{tab:PSC}
\footnotesize
\resizebox{\columnwidth}{!}{%
\scriptsize{
\centering{
\begin{tabular}{p{0.2\columnwidth}|c|c|c|c|c|c|c|c|c|c|c|c|c|c|c|c|c|c|c|c|}
\hline
\textbf{PSC}          & \rot{\shortstack[l]{638.imagick\_s-10316B~~~}} &
\rot{\shortstack[l]{649.fotonik3d\_s-1176B~~~}} &
\rot{\shortstack[l]{607.cactuBSSN\_s-2421B~~~}} &
\rot{\shortstack[l]{625.x264\_s-18B~~~}}& \rot{\shortstack[l]{648.exchange2\_s-1699B}} &
\rot{\shortstack[l]{654.roms\_s-842B}} &
\rot{\shortstack[l]{600.perlbench\_s-210B}} & 
\rot{\shortstack[l]{628.pop2\_s-17B}} &
\rot{\shortstack[l]{644.nab\_s-5853B}} &
\rot{\shortstack[l]{603.bwaves\_s-3699}}& \rot{\shortstack[l]{627.cam4\_s-573B}} &
\rot{\shortstack[l]{621.wrf\_s-575B}} &
\rot{\shortstack[l]{631.deepsjeng\_s-928B}} &
\rot{\shortstack[l]{657.xz\_s-3167B}} &
\rot{\shortstack[l]{602.gcc\_s-734B}} &
\rot{\shortstack[l]{641.leela\_s-800B}}& \rot{\shortstack[l]{623.xalancbmk\_s-700B}} &
\rot{\shortstack[l]{619.lbm\_s-4268B}} &
\rot{\shortstack[l]{605.mcf\_s-665B}} &\rot{\shortstack[l]{620.omnetpp\_s-874B}}
\\ \hline \hline
\shortstack[l]{\textit{nl-mlop-kpcp-nl}} &  & \OK & \OK & \OK & \OK & & &  \OK & \OK & \OK & \OK & & \OK &  & \OK & &  &  & & 

\\ \hline
\shortstack[l]{\textit{nl-bingo-spp-nl}} &	&\OK	&	&	&\OK	& &	\OK&	\OK	&	&	&	&	&\OK	&	&	&\OK	&	&	&	&\OK
\\ \hline
\shortstack[l]{\textit{nl-bingo-kpcp-nl}} & & \OK & & & \OK & & \OK & \OK & & & & & & & & \OK & & \OK & & 
\\ \hline

\shortstack[l]{\textit{no-mlop-kpcp-nl}} & & \OK & & \OK & & & &\OK & \OK &  & \OK &  & \OK & &\OK &  & &  && 
\\ \hline 

\shortstack[l]{\textit{nl-mlop-spp-nl}} & & & \OK & \OK & \OK & & & \OK & \OK & \OK & \OK & & \OK & & \OK & & &  && 
\\ \hline 

\shortstack[l]{\textit{no-bingo-kpcp-nl}} & & \OK & & & & & \OK & \OK & & & & & & &  & \OK &  & \OK & &  \OK
\\ \hline 
\shortstack[l]{\textit{no-bingo-spp-nl}} & & \OK & & & & & & & & & &  & \OK & &  & \OK & \OK &  &  & \OK
\\ \hline 
\shortstack[l]{\textit{no-bingo-kpcp-no}} & & \OK & & & & & & & & & & & & & & & &  & \OK & 
\\ \hline 
\shortstack[l]{\textit{nl-bingo-ipcp-no}} & & & & & & & & & & & & & & & &   & \OK &   & \OK & 
\\ \hline 
\shortstack[l]{\textit{no-bingo-ipcp-no}} & & & & & & & & & & & & & & & &  & \OK & & \OK & 
\\ \hline 
\end{tabular}
}
}
}
\vspace{-0.2in}
\end{table}

{\color{black}
To handle this problem we propose to use only the hardware-invariant events as our features. An example of a hardware-invariant event is the number of conditional branches, which is not affected by the choice of PSC. The use of hardware-invariant events makes it faster to generate the data set required for training \myalgo~and allows us to cover all scenarios during training. 
\ignore{
To find our hardware-invariant features we run the 20 \trace s available in the ChampSim repository \cite{champsim}. We run each trace for 2M instructions.
We collect 180 E-PTI values per \trace at a 1M instruction window size (1 average E-PTI value per instruction window per event per \trace for  different events
}
For generating the training data, we use all available prefetchers in the ChampSim repository as well as the 1st (IPCP \cite{Pakalapati2020ipcp}), 2nd place (Bingo \cite{bakhshalipour2019bingo}), and 3rd (MLOP \cite{shakerinava2019multi}) finalists of the 3rd data prefetching competition (DPC3) \cite{dpc3}. In the table shown in Figure \ref{fig:sysarchitecture}, we show the different prefetchers we used at each memory level. We check the variance of each E-PTI value (for 180 total hardware events) for each PSC. We identify 59 events that are hardware-invariant and have a maximum variance below 10\% from their mean value across all PSCs. We further reduce the number of events by eliminating the redundant events that track similar behavior. 
\ignore{
For example, indirect branches, branch not taken, and branch direct jump all give information regarding the branching behavior of the application phase. So we only consider ?? events.} Table \ref{tab:feature_set} shows the 6 events we use to track trace behavior.

In addition, to reduce the hardware overhead of \myalgo we want to avoid choosing multiple PSCs covering the same traces and we want to reduce the diversity in the prefetchers. To this end, we initially run the 20 traces available in the ChampSim repository with all possible PSCs (2 L1I\$ $\times$ 5 L1D\$ $\times$ 6 L2\$ $\times$ 2 LL\$ =  120 PSCs). For each trace, we sort the PSCs based on the corresponding IPC values in descending order. We then generate a new table for each trace, where the table contains the top 10 PSC entries for the trace. We then combine these tables to form a super-table that contains the top 10 PSCs for all traces. Note that a PSC may be in the top 10 for more than one trace. For our case, we end up with 84 unique PSCs. We sort the PSCs in descending order based on the number of traces for which the PSC is in the top 10. Starting from the top, we select just enough PSCs to improve performance of all the 20 traces. We eventually end up with the 10 PSCs shown in Table \ref{tab:PSC}. Here the tick indicates that the PSC ranks in the top 10 for the corresponding trace. Note that performance of \textit{638.imagick\_s-10316B}, \textit{654.roms\_s-842B}, and \textit{657.xz\_s-3167B} is agnostic of the choice of PSC. It is interesting to note that the two best PSCs from DPC3 -- \textit{no-bingo-nl-nl} and \textit{no-ipcp-ipcp-nl} -- are not in the top 10 choices for PSCs shown in Table~\ref{tab:PSC}. 

\ignore{
We then compare the coverage (the highest performance for the most amount of traces) of the PSCs and choose the 10 PSCs that provide the best coverage for the 20 traces. 
The following three traces are not covered by the 10 PSC in the table because almost all PSCs result in an IPC that is within 0.5\% of the best performance: \textit{638.imagick\_s-10316B}, \textit{654.roms\_s-842B}, and \textit{657.xz\_s-3167B}. }
After we have identified our hardware-invariant events that will be the features and the PSCs that will be the classes of our ML model, we run the $185$ \trace s generated from SPEC2017 (as used by the most recent prefetcher competition) for 120M instructions and collect E-PTI values for the hardware-invariant events using 1M instruction windows. We have a warm up phase of 20M instructions before we start collecting the E-PTI values. In total we have 185(traces)*100(instruction windows per trace) = 18500 instructions windows that make up our dataset. In the dataset we have an IPC value per PSC for each instruction window as our labels. We predict the best PSC for the next instruction window.

In addition to leveraging hardware invariance, to make sure that our algorithm is not overfitting the data, we severely limit the training set size by dividing the dataset into two disjoint sets - 10\% of instruction windows form the training set and the remaining 90\% of the instruction windows form the testing set and perform 10-fold cross-validation on the training set.} 
We form our suite of random forests wherein we train a separate forest for each class (i.e. each PSC) using CART (classification and regression trees) \cite{steinberg2009cart}.

\ignore{
\footnote{CART is a greedy recursive search algorithm that maximizes information by splitting the data at each node using one feature. Each child node is split recursively until there is no information gain from splitting a child node.}} 
We limit the total number of nodes in \myalgo~to keep the size of \myalgo~smaller than L1\$. With this limitation in mind, we conduct a hyper-parameter search and find that the number of estimators (trees per random forest) should be 5 and the number of nodes should be 50 per class. \myalgo~is trained to find the best IPC value for the next instruction window. We would like to note that \myalgo~is made up of several random forests and each forest is made up of several decision trees. This increases the tolerance of our method where even if some of the trees give wrong decisions, other trees can overcome this error.

\ignore{
 \begin{figure}[tb]
   \centering
   \includegraphics[width=\columnwidth]{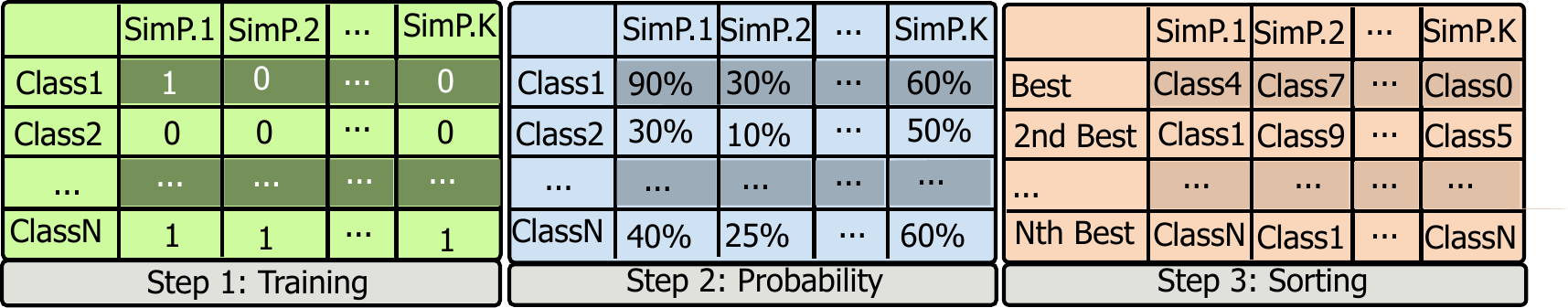}
   \caption{\textbf{Steps of \myalgo~\suitapmech~-} Step 1 determines which
 prefetch
 configurations are suitable for a \trace. Step 2 calculates the
 probability each tree associates with a \trace~belonging to its class. Step 3
sorts prefetch
 configurations based on the probabilities.}
   \label{fig:steps}
 \end{figure}
We will now walk through the training and testing of \myalgo. Figure
\ref{fig:steps} shows the steps of \myalgo. In step 1, we split the oracle
dataset into two datasets: training dataset and test dataset. We further split
the training dataset into 10 validation datasets. For each \trace, in the table,
we assign a ``1'' to the row corresponding to the class that gives the highest
IPC. Here each class corresponds to a prefetch configuration. We also assign
``1'' to rows that given an IPC value within $\%0.5$ IPC of the highest IPC
value.\ignore{\footnote{For a given \trace, when we assign a ``1'' to a prefetch
configuration, this means this prefetch configuration is \textit{suitable} for
this \trace. A ``0'' means this prefetch configuration is \textit{not suitable}
for this \trace.}} Once we have labeled our training \trace s, we construct a
binary tree for each prefetch configuration using the entries in its
corresponding row. In step 2, we perform 10-fold cross-validation on the binary
trees. Using the validation datasets, we get a probability that a \trace~belongs
to a specific class and we can check to see if we made a correct choice. We
repeat steps 1 and 2 with the 10-fold validation datasets to find the best set
of hyperparameters\footnote{The hyperparameters are the maximum depth of a tree,
the minimum elements a leaf must contain, the maximum nodes allocated to a tree,
the maximum performance metrics allowed, and type of accuracy metric (GINI or
entropy).} for training our model. Using IPC as the metric, we evaluate the
impact of each hyperparameter set on the classification. We choose the best set
of hyperparameters after exhaustive search. In step 3, we train our classifier
using the best set of hyperparameters we found and the full training dataset. We
evaluate our classifier using the test dataset. We obtain probability of a test
\trace~ belonging to a class. We end up determining $N_{psc}$ probability values
for each \trace. We sort the classes based on these probability values. When
selecting a prefetch configuration for a \trace~we have two options. In the
first option, we can choose the class with the highest probability obtained for
each \trace. In the second option, we can cycle through the top $N_{out}$
classes.\footnote{As part of our study, we vary $N_{out}$ from one to five.} By
leveraging this output set, \myalgo~has a higher chance of containing the actual
``best'' class. In this work, we focus on the first option, singular output. In
addition to singular output results, we provide a brief study for the second
option, multiple-output (details in Section \ref{sec:multi_out_res}). After we
have a best \myalgo~we add \myalgo~into our cycle-accurate simulator. We run a
given \trace~and make a decision upon the prefetch configuration every
instruction window.}

\subsection{\myalgo~Hardware Design}
\label{sec:hardware}

\begin{figure}[]
  \centering
  \includegraphics[width=0.8\columnwidth]{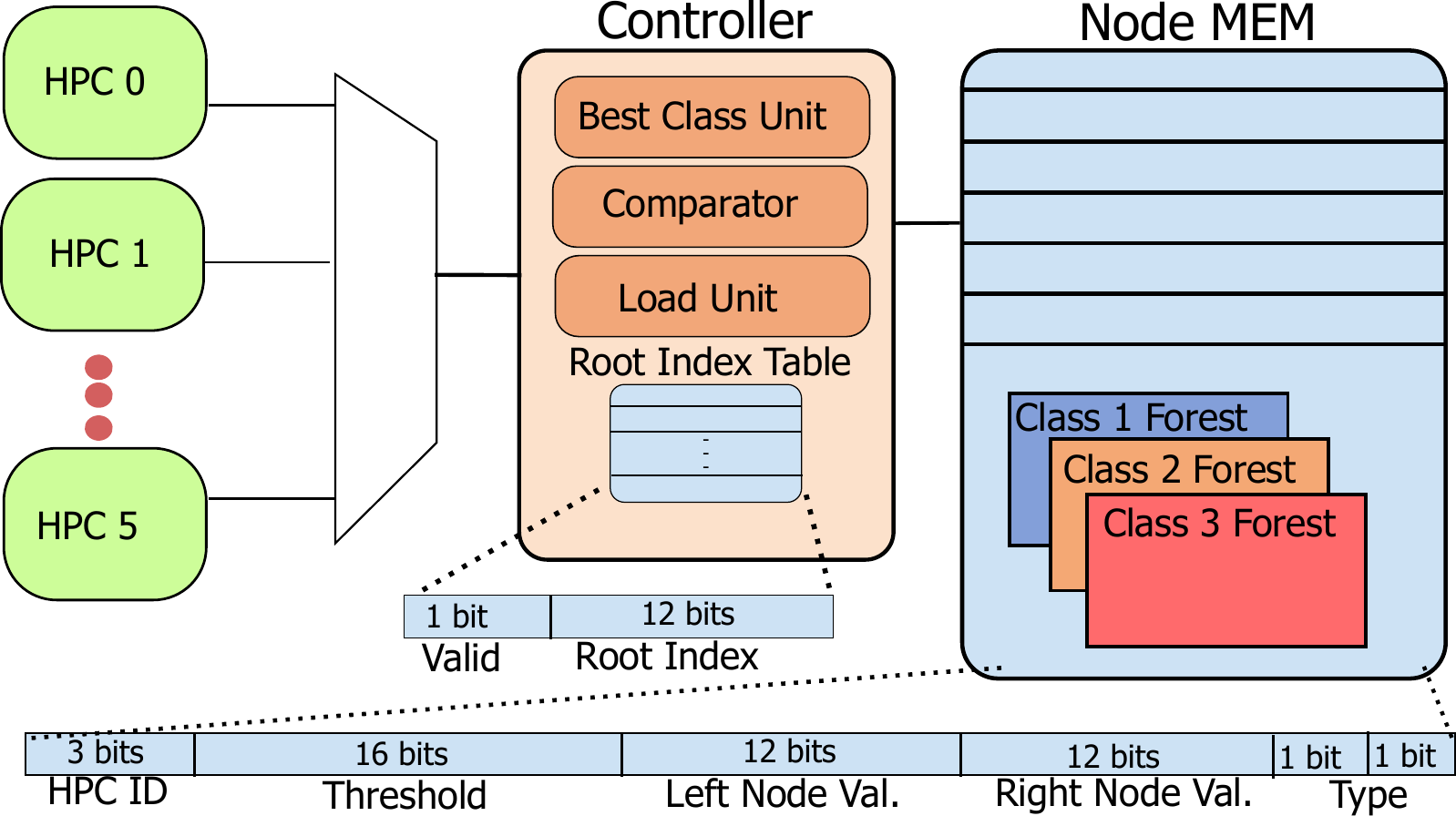}
  \caption{\myalgo~hardware design.}
  \label{fig:structure}
  \vspace{-0.2in}
\end{figure}
\ignore{
We use a total of 2250 nodes, which allows us to store \myalgo~in an SRAM array of a smaller size than L1\$, to design the suite of random forests. We use a single port SRAM array called \textit{Node MEM} to store information about {\color{black}all the 2250 nodes.}} 

Figure \ref{fig:structure} shows the hardware design of \myalgo. We use a single port SRAM array called \textit{Node MEM} to store information about \myalgo. We load the offline-trained model into the \textit{Node MEM} at startup using firmware. Each entry of \textit{Node MEM} corresponds to one node in one of the random forests and it consists of the following fields:
(i) A 3-bit \texttt{HPC ID} field that specifies which hardware-level event is used by that node to make a decision. The 3-bit encoding enables the node to use one of 6 different hardware-level events (see Table~\ref{tab:feature_set}).
(ii) A 16-bit \texttt{Threshold} field (threshold value is determined during training), {\color{black}which is employed {\color{black}by the node} to make a decision if the decision path should branch left or right}.
(iii) A 12-bit (for 2250 node addresses) \texttt{Left~Node~Value(LNV)} field, and (iv) a 12-bit \texttt{Right~Node~Value(RNV)} field. These \texttt{LNV} and \texttt{RNV} fields represent child node indices for internal nodes of a tree. For the leaf nodes of a tree, we use these \texttt{LNV} and \texttt{RNV} fields to indicate the probability of a class.
We differentiate between child node index and probability using (v) a 1-bit \texttt{Type} field.

At the end of every instruction window, \myalgo~calculates the probability of using each class in the next instruction window by the traversing the trees of the associated forest and using E-PTI values for the current window as inputs. For each forest, the controller in \myalgo~reads the \textit{Node MEM} index of the root {\color{black}node} for the first tree from \textit{Root Index Table (RIT)} and loads the \textit{Node MEM} entry for {\color{black}the} root node using a \textit{Load Unit} into a register. Next, the \texttt{HPC} \texttt{ID} in the loaded \textit{Node MEM} entry is used to load the corresponding E-PTI value into a second register. Then the \texttt{Threshold} value, stored in the first register, and E-PTI value stored in the second register are compared using the \textit{Comparator}. Based on the \textit{Comparator} output, we choose the left child or the right child. The \textit{Controller} then uses the {\color{black}corresponding index value from \texttt{LNV} or \texttt{RNV}} to find the next node in \textit{Node MEM}. The \textit{Controller} continues traversing the tree until it loads a probability value corresponding to a leaf from the \textit{Node MEM}. The above steps are repeated for the remaining trees in the forest, and then we calculate the average of the probability values obtained from all the trees in that forest. The \textit{Best Class Unit} in the \textit{Controller} stores the ID of the class with the highest probability value. Every time the \textit{Controller} finishes traversing a forest, the probability value of that forest, i.e. class, is compared with the probability value stored in the \textit{Best Class Unit} using the \textit{Comparator}. If the new probability value is higher than the current value, the \textit{Best Class Unit} updates the probability value and the ID of the class. Once all forests have been traversed i.e. all classes have been evaluated, \myalgo~chooses the entry stored in the \textit{Best Class Unit} as the PSC for the next instruction window.

\ignore{
At the end of an instruction window we do not have to stall the processor until \myalgo~figures out the PSC for the next window.
\myalgo~can determine the best PSC while
instructions are being executed, and we can update the PSC in the background without impacting the on-going execution. This
is possible because when changing the PSC we simply
allow or disallow (gate) outgoing prefetch requests based on whether the
prefetcher is ON or OFF. We do not enable or disable the prefetchers. This
choice also has an added benefit of allowing gated prefetchers to still get
partially trained at runtime and to stay prefetch ready.}

We determined that a maximum depth of $10$ per tree is more than sufficient to accurately determine the best PSC. In our evaluation we use a prefetcher system with $N_{psc}$=10 (given in Table \ref{tab:PSC} and discussed in detail in Section~\ref{sec:results}). {\color{black}If we evaluate all 10 forests in series, where we will require a maximum 500 comparison operations (10 forests * 5 trees per forest * 10 comparisons = 500 comparisons),} it will take less than $0.1\%$ (assuming each comparison operation takes less than a clock cycle) of the time required to execute the 1M instructions in the instruction window. Thus we end up using the chosen PSC for 99.9\% of the instruction window.

\ignore{\footnote{We use memory size as proxy for hardware overhead.}} We need a total of 2250 nodes to design the trees in \myalgo, and these nodes require a {\color{black}12.75} KB-sized \textit{Node MEM} (compared to L1\$ of 64 KB). Other than \textit{Node MEM}, we require a $5*N_{psc}$-entry \textit{RIT} where each entry is 13-bit wide (12 bits for the root node index and 1 valid bit), a 12-bit comparator, a load unit, and a register to store the best class information in the \textit{Controller}.  {\color{black}For the \textit{Node MEM} operating at $5GHz$, using Cacti 7.0 \cite{balasubramonian2017cacti} we find that the area overhead in a $32nm$ process is roughly $0.03mm^2$ and the power is roughly $0.2mW$ for 500 \textit{Node MEM} accesses every 1M instructions. \ignore{These area and power overheads would be lower for more aggressive technology nodes.} The area and power required for the remaining components is negligible.}

\ignore{
Although
training using each one of the $N_{psc}$ * 2051 \trace s or entries from the dataset is
essential for robustness, {\color{black}from our experiments, we discovered that
some entries are not necessary as the application behavior can be captured with
the remaining entries.} As an example, consider a program where the program
accesses data with a constant stride. The only PSC that
might matter for this \trace~is when the Stride prefetcher is switched ON and
when it is OFF. Thus training only using those entries from the dataset where
the Stride prefetcher is ON and where the Stride prefetcher is OFF is
sufficient. For each class (i.e. PSC), we calculate
the percentage of entries from the dataset that show IPC gain. If that
percentage value is below a certain threshold, we do not train for that class
i.e. we prune that class. For example in Table \ref{tab:app_example}, if we set
a pruning threshold to 0.6, only entries corresponding to Class0 will be used to train \myalgo~because
Class0 benefits three out the four benchmarks, i.e., $75\%$ of the benchmarks,
and all other classes benefit less than $60\%$ of the benchmarks.}
\ignore{
To track
program behavior as faithfully as possible, we started with a large number of
hardware-level events ($>$100), and reduced the number of hardware-level
events by eliminating ones that indicate similar behavior. Table
\ref{tab:feature_set} shows the top 15 hardware-level events we found to
be most suitable to track application behavior. We trained using these 15
hardware-level events, and counted the number of times each
hardware-level events was used by \myalgo~and retrained using the top 10
hardware-level events. We repeated this step to further prune the
hardware-level event count to 5.}
\ignore{
Pruning the number of
classes used for training does not restrict \myalgo~from choosing those classes.
This means that we will still need to construct $N_{psc}$ trees. We can prune the
number of constructed trees using the pruning thresholds so that we have minimal
impact on the accuracy of \myalgo. {\color{black}The trees that are pruned do
not have to correspond to the classes that were pruned during training. Pruning
a tree means we can no longer choose this class. We would like to note that pruning a class used for
training does not imply the same thing. Since the tree for that pruned class could still be constructed we can still end up choosing that class. Through tree pruning,} we
can allocate more nodes for the remaining trees and increase their prediction
accuracy.}
\ignore{
Some classes
(i.e. PSCs) can lead to large variance in program behavior. So,
intuitively, these classes require more nodes to achieve high accuracy, while
other classes can use fewer nodes. Allocating nodes based on the requirements of
a class allows us to use hardware resources more efficiently. We judge the quality of training by computing the IPC gain (over the baseline PSC) for each \trace~when using the PSC {\color{black}that we predicted would be best for the \trace.}  We refer to
the average performance gain for each class as $P_{cx}$ for a class $C_x$, where x can take the value from 1 to $N_{psc}$). Auto-optimization
incrementally allocates a few nodes to each tree, i.e. class, in order starting from $x=1$. For every class, we check the change in $P_{cx}$ (i.e. $\Delta(P_{cx})$). If
$\Delta(P_{cx}) <= 0$ (i.e., the extra newly allocated nodes did not increase
the IPC gain of that tree) then these allocated nodes are reallocated to the next
class. If $\Delta(P_{cx}) >= 0$, we start allocating the next small batch of nodes. Once all nodes have been distributed, any tree with its associated $P_{cx} == 0$ is completely pruned. If a tree is pruned, we retrain from scratch to reallocate all nodes, until all valid $P_{cx}>0$.}

\begin{figure}[]
  \centering
    {
  \includegraphics[width=0.95\columnwidth]{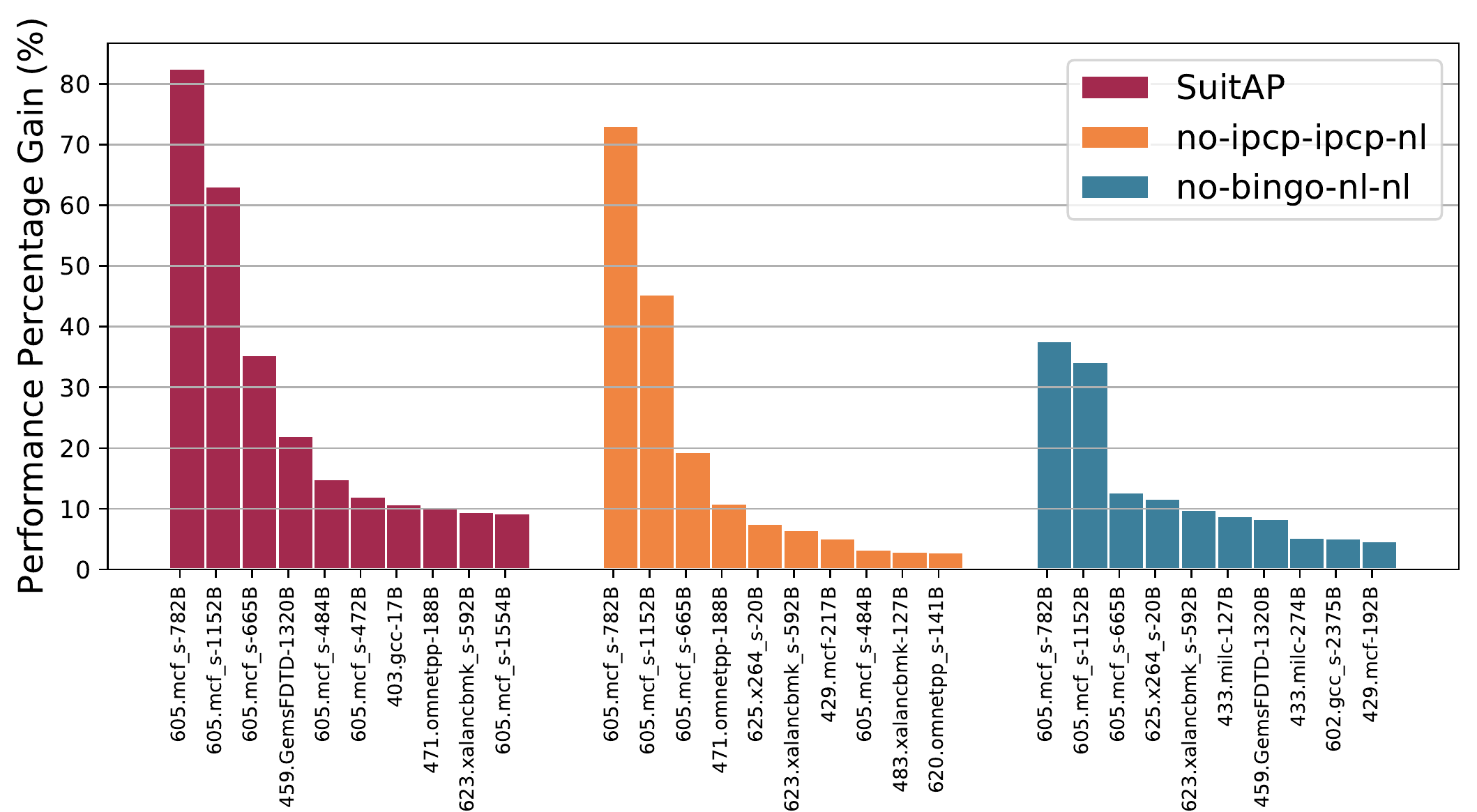}
  }
  \vspace{-0.1in}
  {
  \includegraphics[width=0.95\columnwidth]{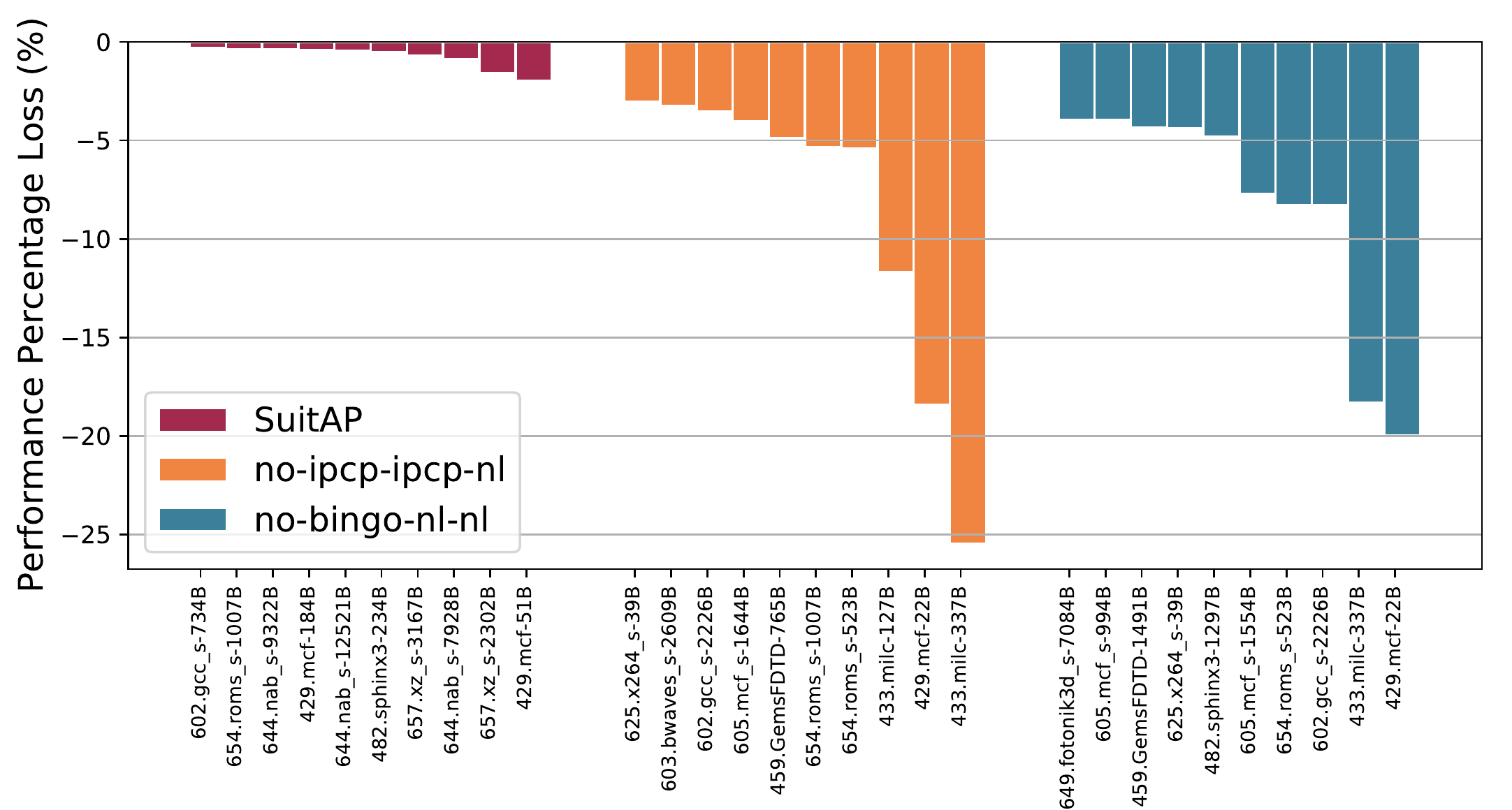}
  }
  \caption{Performance gains (top figure)/losses (bottom figure) of the top 10 SPEC2017 traces when using (i) \textit{no-ipcp-ipcp-nl}, (ii) \textit{no-bingo-nl-nl}, and (iii) \myalgo. Here performance gain or loss is w.r.t. \textit{nl-mlop-kpcp-nl}.}
  \label{fig:spec}
  \vspace{-0.2in}
\end{figure}
\section{Evaluation Methodology}
\label{sec:methodology}
{\color{black}
For our evaluation, we use ChampSim \cite{champsim} to model an OoO processor and multiple prefetchers at each level in the memory hierarchy (see table in Figure~\ref{fig:sysarchitecture}). We use perceptron for branch predictor and least-recently used (LRU) policy for cache replacement policy. We use the default parameters in ChampSim for the rest of the processor architecture.} {We use traces generated from SPEC2017 \cite{SPEC2017} for performance evaluation of \myalgo.}

\ignore{
\begin{table}[t]
  \caption{\textbf{Hardware-level events used by \myalgo -} The ticks
  indicate relevant behavior.}
  \label{tab:feature_set}
\footnotesize
\resizebox{\columnwidth}{!}{%
\scriptsize{
\centering{
\begin{tabular}{p{0.6\columnwidth}|c|c|c|c|c|c|c|c}
\hline
\textbf{Hardware-level events}          & \rot{\shortstack[l]{Performance}} &
\rot{\shortstack[l]{Prefetcher\\Memory \\ Interaction}} &
\rot{\shortstack[l]{Memory\\Access \\ Behavior}} &
\rot{\shortstack[l]{Control \\ Flow}}
& \rot{\shortstack[l]{Memory \\ Hierarchy \\ Interactions}} &
\rot{\shortstack[l]{Temporal\\ Execution \\ Behavior}} &
\rot{\shortstack[l]{Program\\ Execution}} &  \rot{\shortstack[l]{Ranking}}
\\ \hline \hline
\shortstack[l]{Physical Registers Deallocated At Retire} & & & & & &  \OK && 1
\\ \hline
\shortstack[l]{Prefetched L2 Lines Evicted} & & \OK & & & & & & 2      \\
\hline
\shortstack[l]{Stores Commited} & & \OK & & & & &  &3           \\ \hline
\shortstack[l]{Micro-op\$ Hit}     & & & & & &  \OK & & 4       \\ \hline
\shortstack[l]{IPC}  & \OK & & & & & & & 5    \\ \hline
\shortstack[l]{Address Generations For Loads}      & & \OK & & &  & \OK && 6
 \\ \hline
 \shortstack[l]{L1 I\$ tag\$ Hit L1 I\$ Miss}   & & \OK & & & & & & 7
\\
 \hline
 \shortstack[l]{Load Statuses Bad Due To L1 D\$ Misses}  & & & & & \OK &  & & 8
       \\ \hline

\shortstack[l]{L1 I\$ or L1 D\$ Hits} & & & \OK &  & & & & 9         \\ \hline
 \shortstack[l]{L1 D\$ Fills On L1 I\$. Miss}  & & & & & \OK & & & 10
 \\ \hline

 \shortstack[l]{Instr. w. Microcode Retired} & & & & & & &  \OK &11  \\ \hline
 \shortstack[l]{D. Lines Evicted From L2\$} & & \OK & & & & & &  12          \\
 \hline
 \shortstack[l]{L2\$ TLB Reloads} & & \OK & & & &   \OK && 13       \\
 \hline
\shortstack[l]{L1 D\$ Misses}  & & & \OK & & & & & 14        \\ \hline
\shortstack[l]{L1\$ BTB Misses}  & & & &  \OK & & & & 15             \\ \hline
\end{tabular}
}
}
}
\end{table}
}

\section{Evaluation Results}
\label{sec:results}

{\color{black}
Broadly, compared to a processor with no prefetchers, \myalgo~provides an average performance gain of 46\%. For better insight into \myalgo, we also compare the performance of (i) \textit{no-ipcp-ipcp-nl} (DPC3 1st place)~\cite{Pakalapati2020ipcp}, (ii) \textit{no-bingo-nl-nl} (DPC3 2nd place)~\cite{bakhshalipour2019bingo}, and (iii) \myalgo~(we show the prefetchers used at each memory level by \myalgo~in bold in the table found in Figure \ref{fig:sysarchitecture}); against \textit{nl-mlop-kpcp-nl} PSC as our experiments show that if we could choose only one PSC, then \textit{nl-mlop-kpcp-nl} would be best choice for 50\% of the traces. L1I\$ prefetchers were not an option during the DPC3. We compared \textit{no-bingo-nl-nl} with \textit{nl-bingo-nl-nl} and \textit{no-ipcp-ipcp-nl} with \textit{nl-ipcp-ipcp-nl} and did not observe a large performance difference, therefore, we leave \textit{nl-bingo-nl-nl} and \textit{nl-ipcp-ipcp-nl} out of the discussion.} 

{\color{black}In Figure \ref{fig:spec}a we show the top ten performance gains for the two PSCs and \myalgo~to understand the distribution of the performance gains. We cannot show the performance gain for all 185 traces due to space constraints. Using \textit{no-ipcp-ipcp-nl} and \textit{no-bingo-nl-nl}, we observe a maximum performance gain of 73.1\%  and 37.6\%, and an average performance gain (across all 185 traces) of 0.4\% and 0.43\%, respectively. As we can clearly see, \myalgo~beats both competitors in terms of performance gains. \myalgo~has a maximum performance gain of 82.5\% and an average of 2.2\% across all 185 traces.}

{\color{black}
The main advantage of \myalgo~is that it minimizes the negative outliers. In Figure \ref{fig:spec}b we show the ten worst trace performances for the two PSCs and \myalgo. \myalgo~has a performance loss in the range of 2\% to 0.3\%. \textit{no-ipcp-ipcp-nl} and \textit{no-bingo-nl-nl}, have a performance loss in the range of 25.5\% to 3\% and 20\% to 4\%, respectively. This clearly shows that \myalgo~provides us with a win-win situation, whereby we not only see a better average performance gain across all traces but also see a reduction in the performance loss for the outliers.}

\ignore{
\begin{figure}[]
  \centering
  \includegraphics[width=\columnwidth]{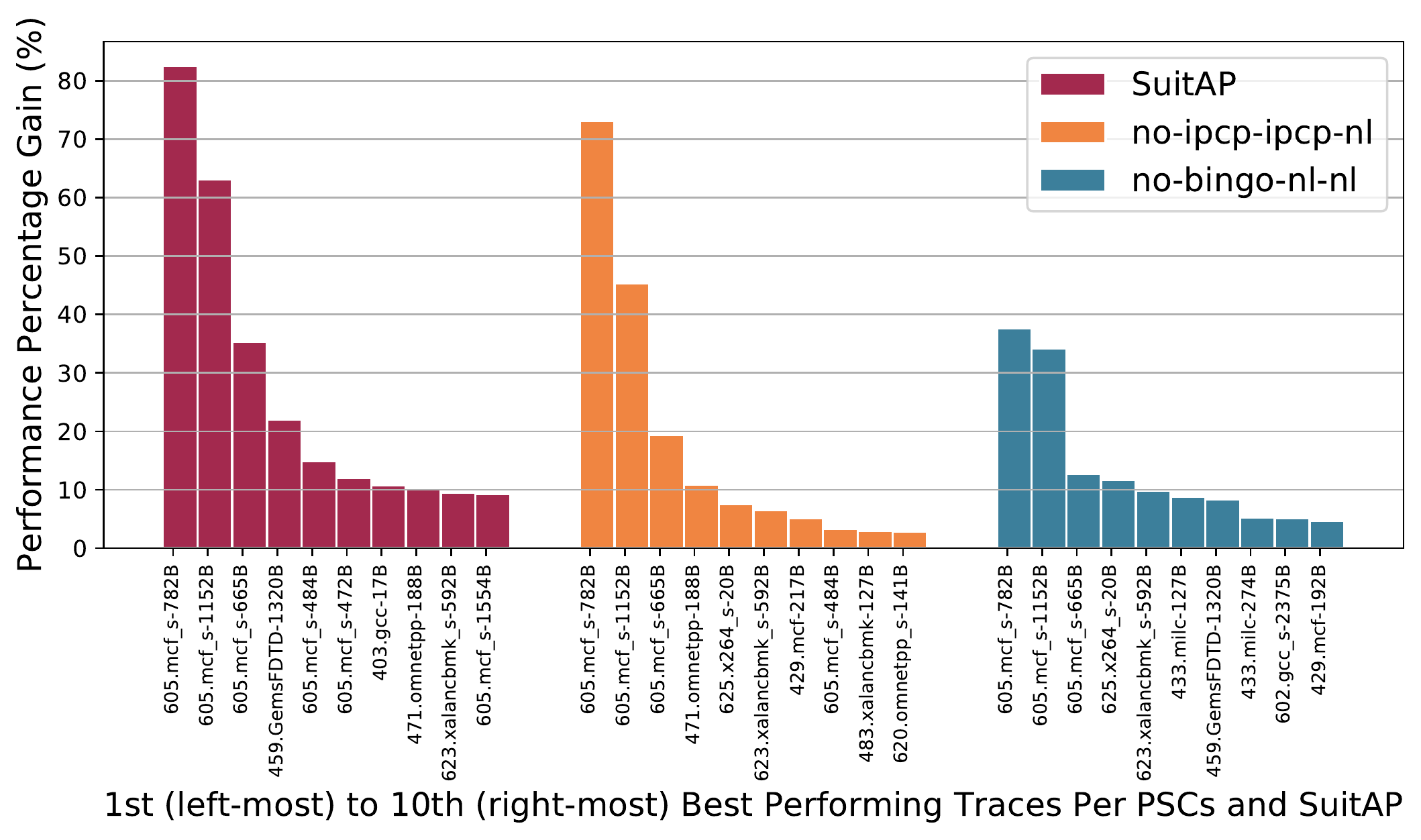} \caption{\textbf{Top Ten Performance Outliers} - Performance normalized to \textit{nl-mlop-kpcp-nl} of SPEC2017 traces when using different prefetchers at each level, and \myalgo~ which chooses a PSC every 1M instructions (see Table~\ref{tab:PSC}) depending on the current trace behavior. Each group signifies the \textit{Nth}-best performance for the two PSC we compare to and \myalgo~ordered 1st-best (left-most) to 10th-best (right-most). Larger is better.}
  \label{fig:spec_pos}
  \vspace{-0.15in}
\end{figure}
}

\ignore{
\begin{figure}[]
  \centering
  \includegraphics[width=\columnwidth]{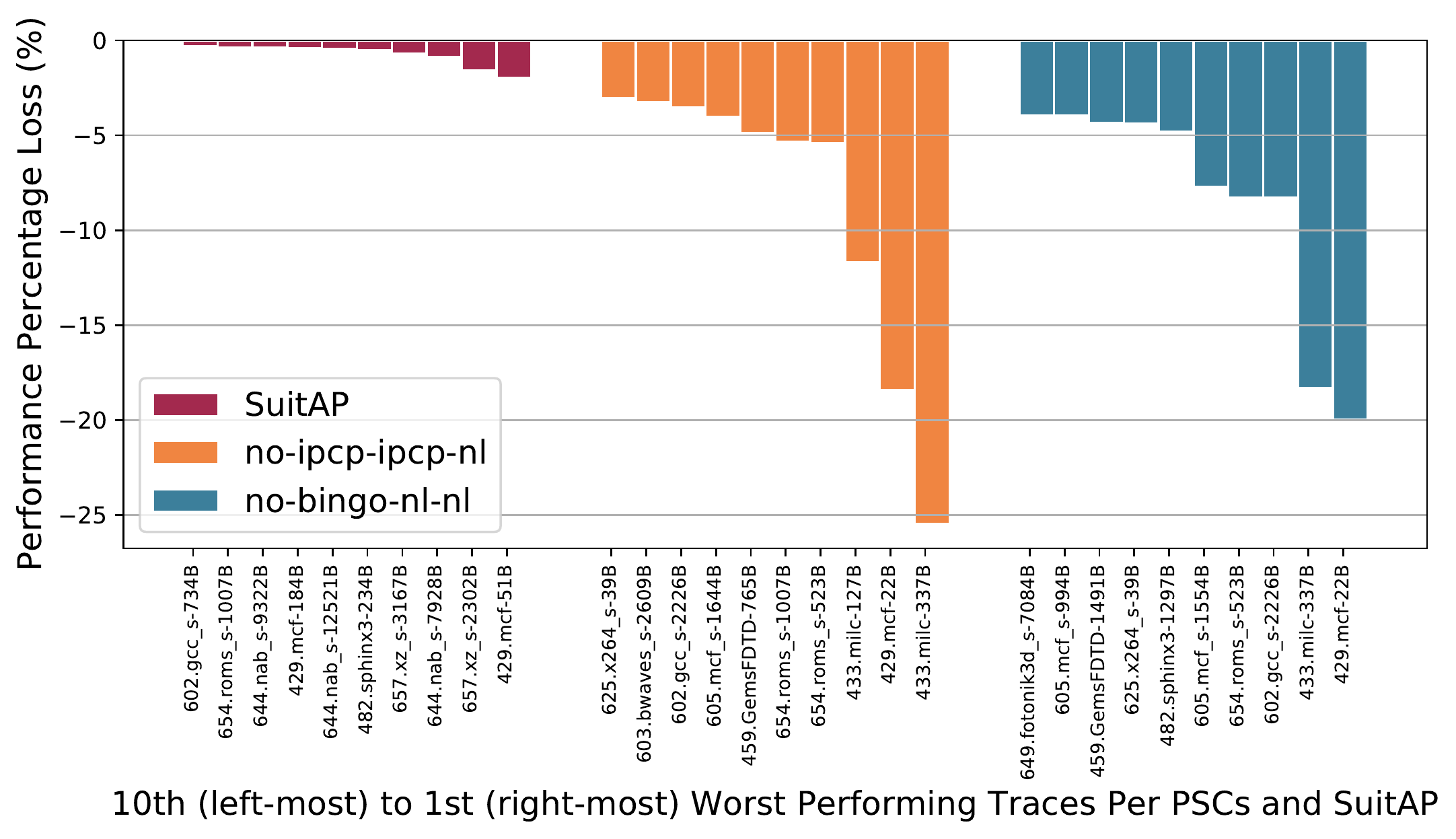}
\caption{\textbf{Bottom Ten Performance Outliers} - Performance normalized to \textit{nl-mlop-kpcp-nl} of SPEC2017 traces when using different prefetchers at each level, and \myalgo~ which chooses a PSC every 1M instructions (see Table~\ref{tab:PSC}) depending on the current trace behavior. Each group signifies the \textit{Nth}-worst performance for the two PSC we compare to and \myalgo~ordered 10th-worst (left-most) to 1st-worst (right-most). Smaller is better.}
  \label{fig:spec_neg}
  \vspace{-0.2in}
\end{figure}
}

\section{Conclusion and Future Work} 
In this work, we introduce \myalgo, a novel ML-based prefetcher adapter designed using custom tailored random forests. We train a dedicated random forest for each PSC, which allows the random forest to retain more information in a smaller amount of hardware. \myalgo~with multiple prefetchers on each level in the memory hierarchy improves the performance of applications by 46\% on average (642\% peak value) when compared to a processor with no prefetching. {\color{black}\myalgo~also reduces the negative outliers from a maximum 25.5\% for ipcp and 20\% for bingo performance loss in the prior work down to a 2\% maximum performance loss.} In the future, we plan to develop \myalgo~for a manycore system where there will be {\color{black}co-ordination between all prefetchers in all of the cores.}

\bibliographystyle{abbrv-etal}
\bibliography{suitap-cal-2020.bib}

\ignore{


\documentclass[9pt,journal,compsoc]{IEEEtran}
\IEEEoverridecommandlockouts
\def\BibTeX{{\rm B\kern-.05em{\sc i\kern-.025em b}\kern-.08em
    T\kern-.1667em\lower.7ex\hbox{E}\kern-.125emX}}


\usepackage{graphics}
\usepackage{textcomp}
\newcommand\sbullet[1][.5]{\mathbin{\vcenter{\hbox{\scalebox{#1}{$\bullet$}}}}}

\usepackage{float}
\usepackage{amssymb}
\usepackage{gensymb}
\usepackage{siunitx}

\usepackage{cite}
\usepackage{amsfonts}

\usepackage[font=small,labelfont=bf]{caption}
\usepackage{multicol}
\usepackage{tabularx}
\usepackage{array}
\graphicspath{{../}}
\usepackage{booktabs}
\usepackage{pifont}
\usepackage{algorithm,algorithmic,amsmath}
\usepackage[section]{placeins} 

\newcolumntype{t}{>{\ttfamily}l}  
\pretolerance=10000
\tolerance=2000
\emergencystretch=10pt
\renewcommand{\baselinestretch}{0.995}
 \raggedbottom
\newcommand*\rot{\rotatebox{90}}
\newcommand*\OK{\ding{51}}
\setlength\tabcolsep{4pt}
\newcommand{\missingcitation}{[{\color{cyan}\textbf{??}}]}
\newcommand{\abs}[1]{\left\lvert#1\right\rvert}
\usepackage{multirow}
\usepackage{array}
\newcolumntype{P}[1]{>{\centering\arraybackslash}m{#1}}
\usepackage{lipsum}

\usepackage{mathptmx} 

\newcommand{\ignore}[1]{}
\newcommand{\sil}[1]{\textcolor{black}{#1}}
\usepackage{fancyhdr}
\usepackage[normalem]{ulem}
\usepackage[hyphens]{url}
\usepackage{microtype}
\usepackage[svgnames,table]{xcolor}


\usepackage[bookmarks=true,breaklinks=true,letterpaper=true,colorlinks,linkcolor=black,citecolor=black,urlcolor=black]{hyperref}


\newcommand{\myalgo}{SuitAP}
\newcommand{\suitapmech}{Algorithm}
\newcommand{\fixme}[1]{\textbf{\textcolor{red}{FIXME: #1}}}
\newcommand{\answer}[1]{\textbf{\textcolor{blue}{ANSWER: #1}}}
\newcommand{\ajay}[1]{\textbf{\textcolor{orange}{AJAY: #1}}}

\newcommand{\trace}{trace}
\newcommand{\tracein}{data point}
\newcommand{\system}{\textit{L1 + L2 Hybrid Prefetching System}}
\newcommand{\systeml}{\textit{L1 Hybrid Prefetching System}}
\newcommand{\treeprune}{Constructed Tree Pruning}
\newcommand{\classprune}{Training Class Pruning}
\newcommand{\featureprune}{Feature Pruning}
\renewcommand{\arraystretch}{1.2}
\def\code#1{\texttt{#1}}

\title{\LARGE \bf Custom Tailored Suite of Random Forests for Prefetcher Adaptation}
\author{}

\begin{document}
\maketitle

\section{Conclusion and Future Work} 
In this work, we introduce \myalgo, a novel ML-based prefetcher adapter designed using custom tailored random forests. We train a dedicated random forest for each PSC, which allows the random forest to retain more information in a smaller amount of hardware. \myalgo~with multiple prefetchers on each level in the memory hierarchy improves the performance of applications by 46\% on average (642\% peak value) when compared to a processor with no prefetching. {\color{black}\myalgo~also reduces the negative outliers from a maximum 25.5\% for ipcp and 20\% for bingo performance loss in the prior work down to a 2\% maximum performance loss.} In the future, we plan to develop \myalgo~for a manycore system where there will be {\color{black}co-ordination between all prefetchers in all of the cores.}





\bibliographystyle{abbrv-etal}
\bibliography{./suitap-cal-2020.bib}


\end{document}}

\end{document}